\documentclass[11pt,twoside]{article}
\usepackage{asp2010}

\resetcounters

\begin{document}

\title{A flexible and modular data reduction library for fibre fed echelle spectrographs}
\author{Danuta Sosnowska$^1$, Christophe Lovis$^1$, Pedro Figueira$^2$, Andrea Modigliani$^3$, Paolo Di Marcantonio$^4$, Denis Megevand$^1$ and Francesco Pepe$^1$}
\affil{$^1$Geneva Observatory, University of Geneva, Geneva, Switzerland}
\affil{$^2$Centro de Astrof'sica, Universidade do Porto, Porto, Portugal}
\affil{$^3$ESO, Garching, Germany}
\affil{$^4$INAF, Osservatorio Astronomico di Trieste, Italy}

\begin{abstract}
Within the ESPRESSO project a new flexible data reduction library is being built. ESPRESSO, the Echelle SPectrograph for Rocky Exoplanets and Stable Spectral Observations is a fiber-fed, high-resolution, cross-dispersed echelle spectrograph. One of its main scientific goals is to search for terrestrial exoplanets using the radial velocity technique. 
A dedicated pipeline is being developed. It is designed to be able to reduce data from different similar spectrographs: not only ESPRESSO, but also HARPS, HARPS-N and possibly others. Instrument specifics are configurable through an input static configuration table. The first written recipes are already tested on HARPS and HARPS-N real data and ESPRESSO simulated data. The final scientific products of the pipeline will be the extracted 1-dim and 2-dim spectra. Using these products the radial velocity of the observed object can be computed with high accuracy.
The library is developed within the standard ESO pipeline environment. It is being written in ANSI C and makes use of the Common Pipeline Library (CPL). It can be used in conjunction with the ESO tools Esorex, Gasgano and Reflex in the usual way.
\end{abstract}

\section{Introduction}

The Data Reduction Library developed for ESPRESSO is designed to be flexible and modular. The idea is to be able to reduce the data from different echelle fibre-fed spectrographs using the same algorithms. The library needs only the instrument configuration table to define instrument specific characteristics and CCD geometry parameters provided in a separate FITS table, to reduce properly the data from a given spectrograph. Thus, each recipe depends on a small number of parameters, controlling the relevant data reduction steps. For now the library is tested on the real data from two spectrographs: HARPS and HARPS North and on the simulated ESPRESSO data. In order to reduce data of an additional spectrograph, its CCD geometry and instrument parameters have to be provided. No other changes in the code are needed.
The ESPRESSO pipeline follows the standard data reduction chain, as presented on Figure 1.

\section{Data Reduction Library}

The top-level part of the ESPRESSO DRL is made of a set of recipes coded in C and using CPL objects, which are able to process the raw data of a given type. The ESPRESSO DRL will provide 11 calibration recipes and a science recipe. Executing these recipes according to the data reduction cascade shown on Figure 1, it will be possible to fully reduce all ESPRESSO calibrations and science data. Parameters of each instrument mode are defined in the configuration file, therefore the corresponding data can be reduced by the same recipe, without any code changes. Each recipe will be executed using either ESORex, Gasgano or Reflex.

\subsection{Overview of Reduction Cascade}
We give below an overview of the global reduction cascade, starting from basic calibrations up to science reduction, like shown on Figure 1.
\begin{itemize}
\item The detector bias level and readout noise are measured on stacked BIAS frames (master bias). The overscan zone is checked whether it can be used to reliably determine the bias level. If so, the standard way to subtract the bias on all kinds of frames will be to use the overscan. If not, the master bias will be use instead.
\item DARK frames are used to measure the average dark current and determine the location of hot pixels. A master dark will be created for QC purpose. It is not foreseen to be used in the following data reduction steps since the dark current is expected to be extremely low (~1 e-/pixel/hour) and dark frames will usually be readout noise limited. 
\item A bad pixel map is obtained from the analysis of LED\_FF frames taken with two different exposure times (linearity check).  LED\_FF images are obtained by direct illumination of the CCD with a LED. The gain is also measured for QC purpose, from the relation between flux level and standard deviation of pixel values. 
\item The linearity curve of the CCD is obtained from the analysis of a sequence of DET\_LIN frames, which are Fabry-Perot (FP) frames with different exposure times. They are preferred over LED\_FF frames to measure the detector linearity because it more closely simulates the global flux distribution in the frames expected for a science exposure.
\item ORDERDEF frames (one per fibre) are used to identify and trace spectral orders and slices on the detector. Only the order/slice center is relevant at this stage. 
\item For each fibre, the order profile in cross-dispersion direction is found using high-SNR, co-added FLAT frames. Then, the orders are extracted from the FLAT frames using this profile, and the spectral flat-field is generated together with the associated error and bad pixel information. The (extracted) blaze function is obtained through smoothing of the FLAT spectra and correction for the spectral energy distribution of the calibration lamp.  
\item The physical size of CCD pixels along spectral orders is calibrated using PIXEL\_ GEOM frames, which consist of a sequence of laser comb frames scanning the pixels by varying the comb parameters. The conversion table between pixel number and physical distance along spectral orders is obtained in this way. 
\item CONTAM frames are used to measure the cross-fibre contamination on fibre A from the simultaneous reference on fibre B (ThAr lamp, laser comb or Fabry-Perot). Contamination frames are used during extraction in the science reduction. 
\item The relative efficiency of channels A and B as a function of wavelength is measured using EFF\_AB frames, which are obtained through blue sky observations. This calibration product is used during sky subtraction in the science reduction. 
\item The wavelength calibration for both fibres is determined using WAVE frames. These can be of different sub-types depending on the source used (ThAr lamp, laser comb, Fabry-Perot etalon). Different algorithms are used to reduce frames from different sources. On fibre A (science fibre), the wavelength solution accuracy is particularly important since it is used to calibrate the science spectrum and therefore sets the radial velocity zero point. On fibre B (comparison fibre), the spectrum mainly serves as drift reference spectrum for the drift measurement on science frames with the simultaneous reference. 
\item FLUX\_CALIB frames are used to compute the absolute efficiency of the instrument using spectrophotometric standard stars. The instrument response curve is used in the science reduction to calibrate the science spectrum in flux. ESPRESSO being a fibre-fed instrument with a fixed sky aperture of 1.0 arcsec or less, the precision of the flux calibration is expected to be low because of highly variable slit losses. 
\item Finally, SCIENCE frames are of different sub-types depending on 1) whether simultaneous reference or sky is present on fibre B, and 2) what reference source is used if applicable. Science reduction makes use of all calibration products listed above and generates extracted S2D spectra and merged, rebinned S1D spectra, together with S2D and S1D error and quality maps. Finally, the CCF of the S2D spectrum is computed and the radial velocity is obtained from a Gaussian fit to the CCF.
\end{itemize}

\section{Current DRS development status}
Three DRS releases were provided up to now. The first one, issued in November 2013, contained two recipes: the reduction of a set of BIAS raw frames (\emph{espdr\_mbias}) and the reduction of a set of DARK raw frames (\emph{espdr\_mdark}). These first two recipes allow to monitor the quality of the detector and to correct for the bias and dark current levels. The elimination of cosmics is treated by applying the sigma clipping algorithm to the stack of minimum 5 frames. The products of these two recipes are: the master bias frame and the hot pixels map. The second release, issued in March 2014 contained the recipe to reduce LED flat frames (\emph{espdr\_led\_ff}). LED\_FF recipe products are: the bad pixels map and the computed detector gain.
Since the ESPRESSO DRS was designed to be able to reduce data from other spectrographs, like HARPS and HARPS North, a more accurate definition of the CCD geometry configuration was needed and this led to the third release, issued by the end of April 2014. We are currently preparing a new release including the recipe to trace the position of the orders on the CCD (\emph{espdr\_orderdef}), to be released end of September 2014.

The data reduced by the ESPRESSO DRS can be analysed further by the Data Analysis Software (DAS), presented in \cite{cupani_2014}.

\includegraphics[width=1.0\columnwidth]{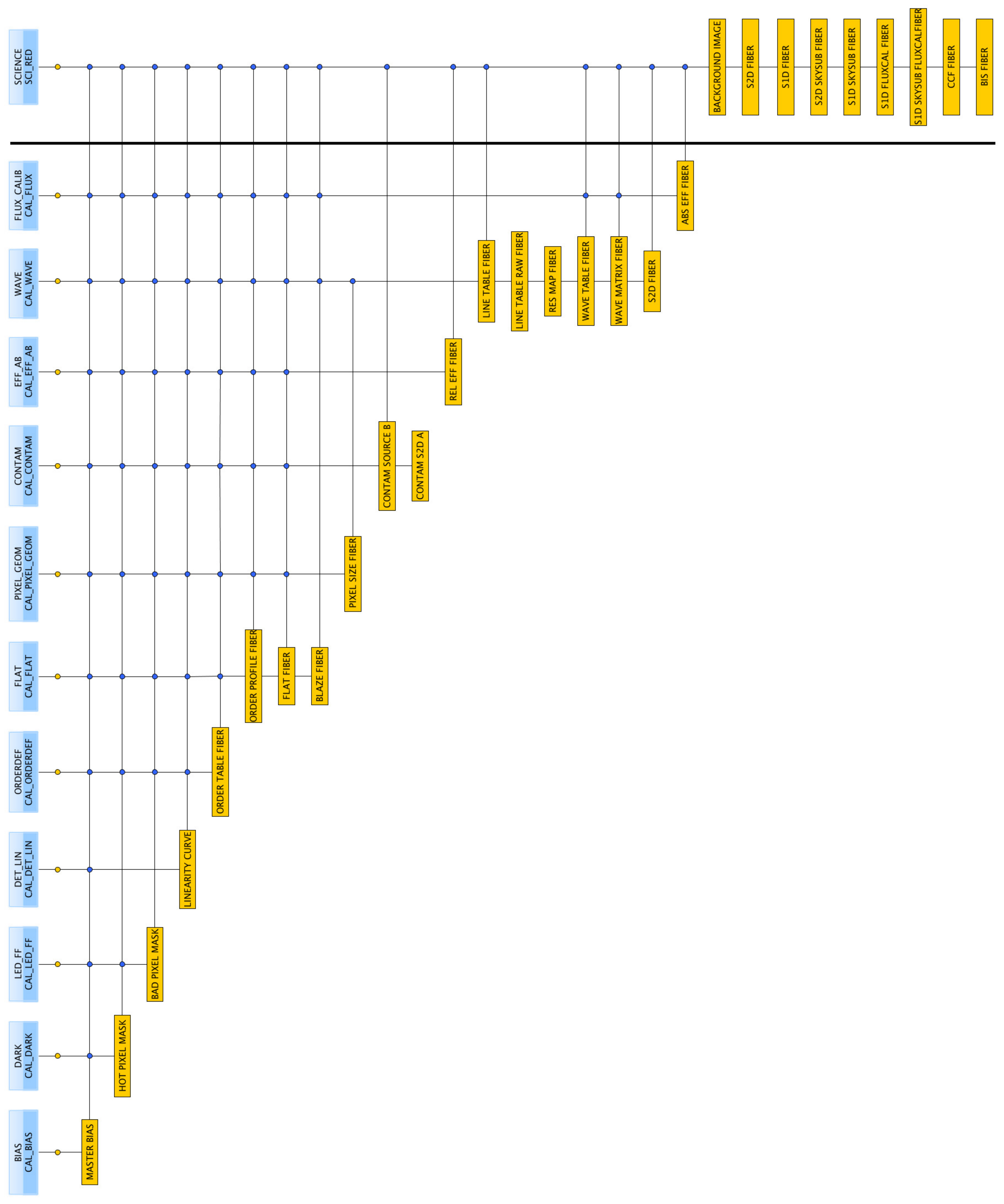}
\textbf{Figure 1} The cascade of the ESPRESSO pipeline recipes


\bibliography{P2-3.bib}

\begin{thebibliography}{}
\expandafter\ifx\csname natexlab\endcsname\relax\def\natexlab#1{#1}\fi
\expandafter\ifx\csname url\endcsname\relax
  \def\url#1{\texttt{#1}}\fi
\expandafter\ifx\csname urlprefix\endcsname\relax\def\urlprefix{URL }\fi
\providecommand{\eprint}[2][]{\url{#2}}

\bibitem[{Cupani et~al.(2014)Cupani, D'Odorico, Cristiani, Gonzalez-Hernandez,
  Lovis, Sousa, Vanzella, Di~Marcantonio, \& Megevand}]{cupani_2014}
Cupani, G., D'Odorico, V., Cristiani, S., Gonzalez-Hernandez, J., Lovis, C.,
  Sousa, S., Vanzella, E., Di~Marcantonio, P., \& Megevand, D. 2014, in ADASS
  XXIV, edited by A.~R. Taylor, \& J.~M. Stil (ASP), vol. TBD, TBD

\end{thebibliography}

\end{document}